\documentclass[conference]{IEEEtran}
\IEEEoverridecommandlockouts

\usepackage{cite}
\usepackage{amsmath,amssymb,amsfonts}
\usepackage{algorithmic}
\usepackage{graphicx}
\usepackage{textcomp}
\usepackage{xcolor}
\usepackage{hyperref}
\usepackage{booktabs}
\usepackage{listings}
\usepackage[table]{xcolor}
\usepackage[table]{xcolor}
\usepackage{listings}
\usepackage{titletoc}
\usepackage{graphicx}
\usepackage[noabbrev,capitalise]{cleveref}

\usepackage{amssymb}
\usepackage{multicol}
\usepackage{multirow}
\usepackage{xspace}
\usepackage{makecell}
\usepackage{tabularx}
\usepackage{bm}
\usepackage{subfigure}
\usepackage{tikz}
\usepackage{tcolorbox}
\usepackage{enumitem}
\usepackage{booktabs}
\usepackage{balance}

\newcommand{\eg}{\emph{e.g.,}\xspace}

\newcommand{\ignore}[1]{}
\newcommand{\paratitle}[1]{\vspace{1.5ex}\noindent\textbf{#1}}

\def\BibTeX{{\rm B\kern-.05em{\sc i\kern-.025em b}\kern-.08em
    T\kern-.1667em\lower.7ex\hbox{E}\kern-.125emX}}
\newcommand{\qiang}[1]{}
\begin{document}

\title{Deep Interest Mining for Intent-Enriched Semantic IDs in Multimodal Generative Recommendation}

\author{
\IEEEauthorblockN{Yangchen Zeng, and Jinze Wang*\footnote{correspondence}}

}

\maketitle
\begin{abstract}
Semantic IDs (SIDs) provide the discrete item vocabulary used by generative recommendation, but their quality depends on what item evidence is preserved before quantization. In product recommendation, surface metadata often misses latent usage intent, visual evidence may be only weakly reflected in text, and downstream policy learning provides sparse feedback about whether a generated SID corresponds to a semantically useful item. We introduce \textbf{DeepInterestGR}, an intent-enriched SID framework for generative recommendation. Before SID quantization, \textbf{CMSA} enriches item representations through two complementary evidence paths: recommendation-oriented VLM captions and projected image embeddings. \textbf{DCIM} then uses an LLM to mine item-side intent descriptors---latent usage motivations implied by product content rather than personalized user states. During policy training over the constructed SIDs, \textbf{QARM} adds a relevance-gated semantic-quality bonus on top of standard SID rewards, applying the bonus only when the generated SID decodes to the target item. Thus, semantic quality cannot reward a fluent but irrelevant item prediction. Experiments on three Amazon Product Review categories (Beauty, Sports, and Instruments) show that DeepInterestGR improves over competitive generative and RL-based baselines, with relative gains of up to \textbf{15.1\%} in NDCG@5 and \textbf{13.9\%} in NDCG@10 over the strongest per-metric baseline. Component ablations, CMSA branch analyses, reward variants, and SID-level case studies support a bounded claim: enriching pre-quantization item evidence with visual cues and item-side intent descriptors, together with relevance-gated semantic rewards, improves SID-based generative recommendation under the evaluated settings. The anonymized code repository provides the implementation, prompts, hyperparameters, encoder settings, random seeds, and computing-resource details: \textcolor{black}{https://anonymous.4open.science/r/DeepInterestGR}.
\end{abstract}


\maketitle

\section{Introduction}

Driven by the tremendous success of large language models (LLMs) across diverse domains \cite{achiam2023gpt,touvron2023llama,zhou2024large}, recommender systems have increasingly shifted toward generative modeling \cite{li2024survey,zhai2024actions,wang2025we}. In contrast to conventional deep learning-based recommenders that rely on multi-stage cascades or funnel-style pipelines, generative recommendation (GR) casts recommendation as next-token prediction and directly generates the next item a user is likely to interact with \cite{zhang2025gpr,zhou2025onerec,han2025mtgr}. This formulation has demonstrated strong empirical performance in real-world applications, including e-commerce recommendation \cite{yi2025recgpt}, search recommendation \cite{wang2025nezha}, advertising \cite{zhang2025gpr}, and video recommendation \cite{zhou2025onerec2}, thereby providing a unified and scalable approach to sequential user modeling.

\begin{figure}[t]
\centering
\includegraphics[width=0.48\textwidth]{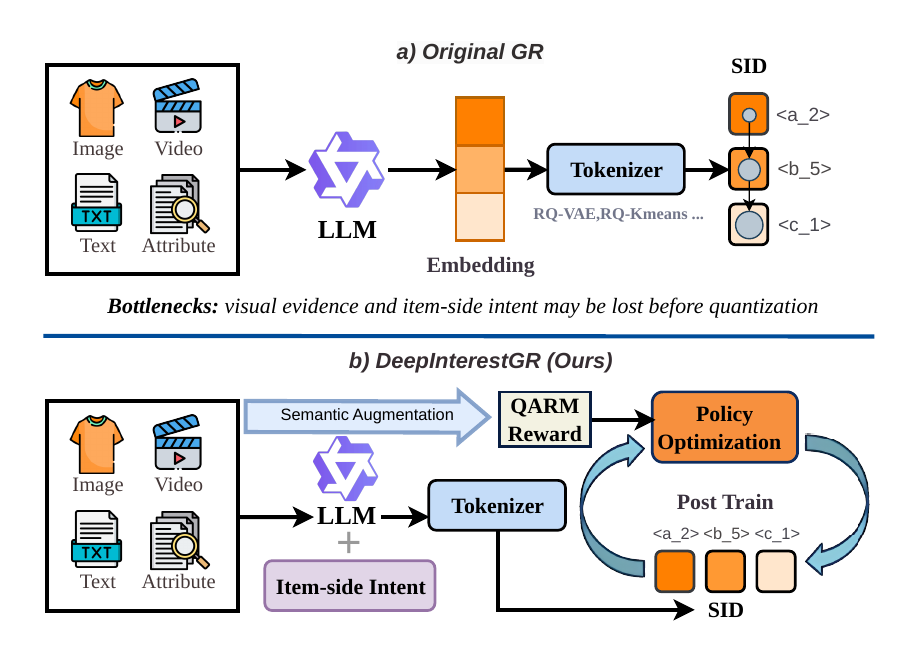}
\caption{High-level comparison between standard SID-based generative recommendation and DeepInterestGR. Standard pipelines compress item content into semantic IDs before generation, which may discard visual evidence and item-side usage intent. DeepInterestGR enriches the pre-quantization item representation with visual evidence, item-side intent descriptors, and relevance-gated policy optimization over generated SIDs.}
\label{fig_motivation}
\end{figure}

Semantic IDs (SIDs) are a key enabler of GR, mapping large item catalogs into compact sequences of discrete tokens \cite{hou2023learningsid}. By compressing the item space while preserving compatibility with next-token prediction, SIDs improve the efficiency and scalability of GR \cite{rajput2023recommender,li2025survey}. In multimodal scenarios, however, an SID is not only an index. It also defines what item semantics are exposed to the autoregressive model. If relevant visual evidence or usage intent is absent before quantization, the resulting discrete code cannot recover it later.

As shown in Figure~\ref{fig_motivation}, many SID generation methods still follow a compression-first design: items are encoded into dense representations and then discretized into token sequences \cite{rajput2023recommender,zhou2025onerec,zhang2025gpr,ye2025align,hou2025generating}. Recent systems such as OneRec \cite{zhou2025onerec}, GPR \cite{zhang2025gpr}, and MiniOneRec \cite{kong2025minionerec} improve this pipeline with stronger generative training or more integrated optimization. Our goal is complementary rather than a claim that all prior methods are strictly two-stage: we focus on the semantic content of the representation that is quantized into SIDs. We identify three recurring bottlenecks:

\paratitle{Weak visual preservation.} Product images often carry recommendation-relevant evidence about style, usage scene, material, or target audience. A caption alone may omit fine-grained visual details, while an image embedding alone may be hard for the text-oriented SID pipeline to interpret. If visual semantics and continuous visual features are not introduced before quantization, the discrete code may underrepresent visual cues that matter for user choice.

\paratitle{Surface-only item semantics.} Titles and descriptions usually state observable attributes, but they rarely express why an item is useful to a user. Surface-similar metadata can hide different usage intents: for example, soccer balls may share generic ``training'' titles while their images indicate youth backyard practice, adult match preparation, indoor futsal control, or outdoor team drills. Text-only SID construction can preserve the coarse product category but still collapse these fine-grained usage intents into nearby prefixes when they are implied by images rather than titles. We call these item-side intent descriptors rather than personalized user interests, because they are inferred from item content and can later support user-history matching.

\paratitle{Sparse policy feedback for semantic quality.} Standard RL or preference-optimization baselines for generative recommendation, including SID-format and binary matching rewards, mainly supervise whether the predicted SID matches the target. They do not explicitly distinguish an SID whose associated descriptor is specific and actionable from one produced from vague or hallucinated semantic expansion.

Based on these observations, we propose \textbf{DeepInterestGR}. The central goal is to enrich item representations before SID quantization so that generated SIDs preserve coarse product categories while separating fine-grained usage intents that are often implied by visual evidence but omitted from surface text. To preserve visual evidence before quantization, \textbf{CMSA} uses a dual-path design: a VLM caption path produces recommendation-oriented visual descriptions, while a visual embedding path projects image features into the same pre-quantization representation used by the SID tokenizer. We use the term alignment in a modest sense: CMSA coordinates visual text, item text, and image embeddings before RQ-VAE, rather than claiming to learn a new contrastive multimodal embedding space. To enrich surface attributes, \textbf{DCIM} mines item-side intent descriptors from the textualized item content. To improve posterior policy learning, \textbf{QARM} adds an LLM-derived semantic-quality signal to RL training, but gates this signal by exact target-item decoding so that semantic quality cannot reward an incorrect prediction by itself.

\paratitle{Contributions.}
We evaluate DeepInterestGR on three Amazon Product Review categories and compare it with traditional sequential models, generative SID models, and RL/preference-optimized recommendation baselines. The main contributions are summarized as follows:

\begin{itemize}[leftmargin=*]
    \item We formulate intent-enriched SID construction for multimodal generative recommendation: before discretization, item representations should preserve visual evidence and item-side usage intent, not only surface product attributes.

    \item We introduce \textbf{DCIM (Deep Contextual Interest Mining)}, which mines item-side intent descriptors from textualized item content. We explicitly distinguish these descriptors from personalized user states: they are item annotations designed to make SIDs easier for the user-history model to match.

    \item We propose \textbf{CMSA (Cross-Modal Semantic Augmentation)}, a dual-path visual enrichment module that injects image-derived evidence before SID quantization through both VLM-generated visual text and projected image embeddings. This preserves interpretable visual semantics while retaining continuous visual features under the same MiniOneRec-compatible tokenizer setting.

    \item We design \textbf{QARM (Quality-Aware Reinforcement Mechanism)}, a relevance-gated RL reward that augments standard SID-format and accuracy rewards with LLM-estimated descriptor quality only when the generated SID decodes to the target item. This avoids rewarding semantically polished but irrelevant items.

    \item Experiments, branch-level CMSA ablations, reward analyses, and failure-oriented SID cases show consistent gains over strong baselines and examine whether intent enrichment changes the generated SID structure, with all reported improvements computed against the strongest baseline for each metric in \Cref{tab:main}.
\end{itemize}

\section{Related Work}

\textbf{Semantic ID for Recommendation.}
Representing items as compact discrete token sequences---commonly referred to as Semantic IDs (SIDs)---has become a foundational technique for scaling recommendation systems. Early approaches constructed SIDs through clustering or hashing over item embeddings and used them as indexing units within retrieval pipelines~\cite{petrov2024recjpq,hou2023learningsid,wang2025hyperman}, treating SID construction as an offline preprocessing step decoupled from the downstream objective. As generative modeling gained traction, SIDs were repurposed as autoregressive generation targets, enabling more expressive item representations~\cite{zhou2025onerec，wang2026meta,zhang2025gpr,li2025survey}. Among quantization strategies, residual quantization (RQ) has emerged as the dominant paradigm: RQ-VAE~\cite{rajput2023recommender}, RQ-KMeans~\cite{zhang2025gpr}, and RQ-KMeans+~\cite{zhang2025gpr} all progressively quantize residual vectors into multi-level codebooks, capturing coarse-to-fine semantic structure. Despite their effectiveness, these methods often leave the semantic content of the quantized item representation under-specified: the model may learn to generate SID tokens accurately while the pre-quantization representation still lacks visual evidence or item-side intent cues. Our work addresses this gap by enriching the representation before quantization and by adding a relevance-gated semantic-quality signal during downstream policy optimization. We do not claim that the RL stage reconstructs the codebook itself; rather, it shapes how the generative policy uses the constructed SIDs.

\textbf{Generative Recommendation.}
The success of large language models in sequence modeling has catalyzed a paradigm shift in recommendation research, from discriminative scoring toward end-to-end generative formulations~\cite{li2024survey,wang2025generative,wang2026agent4poi}. One line of work adapts Transformer-style architectures with novel feature construction schemes to improve generation capacity~\cite{zhai2024actions,han2025mtgr,chai2025longer,zhang2025onetrans,huang2025genrank}. A complementary direction leverages LLMs as offline feature generators or auxiliary signal providers to progressively enhance traditional pipelines~\cite{chen2024hllm,yan2025lum,yi2025recgpt}, without fully replacing the underlying retrieval--ranking stack. However, both directions often retain DLRM-style features or multi-stage cascading designs, which introduce objective misalignment and information bottlenecks~\cite{yan2025lum}. More recent efforts pursue unified end-to-end frameworks that cast user understanding and item generation as a single next-token prediction task~\cite{zhou2025onerec,zeng2026learning,zhang2025gpr}, demonstrating the potential to replace conventional retrieval--ranking pipelines entirely. Our framework builds on this trend while further addressing the semantic quality of SIDs through deep interest mining and reinforcement-based optimization.


\section{Method}\label{sec:method}

In this section, we present the \textbf{DeepInterestGR} framework. It enriches the item representation before SID quantization and then trains the generative policy with relevance-gated semantic rewards. The framework has three components: (1) \textbf{CMSA} (\Cref{sec:cmsa}) injects image-derived evidence through both VLM-generated visual text and projected image embeddings before RQ-VAE quantization; (2) \textbf{DCIM} (\Cref{sec:dcim}) mines item-side intent descriptors from the textualized content; and (3) \textbf{QARM} (\Cref{sec:qarm}) improves downstream SID generation without updating the constructed codebook. All LLM/VLM-derived captions, descriptors, and quality labels are precomputed offline; online recommendation uses only the constructed SIDs.

\begin{figure*}[t]
\centering
\includegraphics[width=\textwidth]{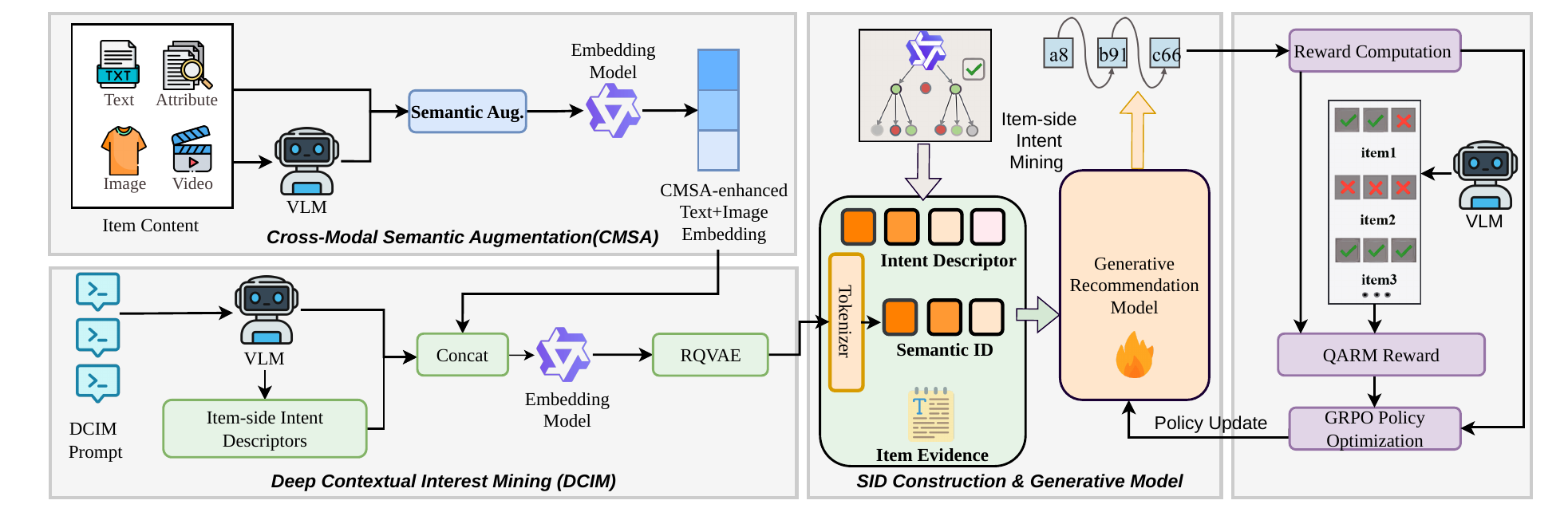}
\caption{Architecture of the DeepInterestGR framework. CMSA enriches item representations before RQ-VAE quantization through dual visual evidence, including VLM-generated visual text and projected image embeddings. DCIM extracts item-side intent descriptors from textualized item evidence. QARM then applies relevance-gated quality-aware rewards during GRPO policy optimization over generated SID sequences, without modifying the constructed SID codebook.}
\label{fig_framework}
\end{figure*}

\subsection{Problem Formulation}\label{sec:problem}

We consider sequential recommendation: given user $u$'s interaction history $\mathcal{S}_u = [i_1, \ldots, i_T]$, predict the next item $i_{T+1}$. Each item $i \in \mathcal{I}$ carries textual metadata $(\mathbf{t}_i, \mathbf{d}_i)$ and visual content $\mathbf{v}_i$. We introduce an item-side intent descriptor $\mathbf{z}_i = \text{DCIM}(\mathbf{t}_i, \mathbf{d}_i, \mathbf{v}_i^{\text{text}})$, where $\mathbf{v}_i^{\text{text}}$ is produced by the CMSA caption path. The descriptor summarizes latent usage motivations implied by item content; it does not use future target information or personalize the item representation to a particular test user. Items are encoded as Semantic IDs via residual quantization~\cite{rajput2023recommender,zheng2024lcrec}: $\mathbf{s}_i = \text{RQ-VAE}(\mathbf{e}_i^{\text{mm}}) = (s_i^{(1)}, \ldots, s_i^{(H)})$, where $\mathbf{e}_i^{\text{mm}}$ fuses item attributes, visual textual evidence, projected image embeddings, and mined intent descriptors.

\subsection{Residual Quantization for SID Construction}\label{sec:gr_framework}

We adopt RQ-VAE to quantize item embeddings $\mathbf{e}_i \in \mathbb{R}^d$ into hierarchical SID tokens. The quantization iteratively assigns codebook entries: $s_i^{(h)} = \arg\min_{k} \| \mathbf{R}_i^{(h)} - \mathbf{c}_k^{(h)} \|_2$, where $\mathbf{R}_i^{(1)} = \mathbf{e}_i$ and residuals are computed as $\mathbf{R}_i^{(h+1)} = \mathbf{R}_i^{(h)} - \mathbf{c}_{s_i^{(h)}}^{(h)}$. The final SID $\mathbf{s}_i = (s_i^{(1)}, \ldots, s_i^{(H)})$ captures item semantics at multiple granularity levels, where each token (\eg \texttt{a8}, \texttt{b91}) corresponds to a discrete index in the hierarchical codebook. Following the MiniOneRec-compatible setting, we use $H=3$ in our experiments.

\subsection{Deep Contextual Interest Mining (DCIM)}\label{sec:dcim}

The core innovation of our framework lies in \textbf{Deep Contextual Interest Mining (DCIM)}, as illustrated in Figure~\ref{fig_framework}. DCIM extracts item-side intent descriptors from item metadata and visual textual evidence. We use the word ``interest'' in the recommender-system sense of latent purchase or usage motivation, but the descriptor is item-side: it is inferred from item content and then used by the sequential model to match user histories. This distinction avoids target leakage and clarifies that DCIM is not generating a personalized profile for each test query.

\paratitle{Motivation.}
Existing SID generation methods often encode items using shallow textual features (\eg product titles and descriptions), which capture explicit attributes but fail to reveal the latent motivations that explain why users interact with the item. For example, given a product ``noise-canceling headphones'', a shallow encoder captures surface attributes such as brand and specifications, but misses intent descriptors such as ``focus-oriented work style'', ``frequent traveler'', or ``audio quality enthusiast''. We hypothesize that instruction-tuned LLMs can help infer these item-side intent descriptors from metadata and visual textual evidence, thereby enriching the semantic content encoded into SIDs.

\paratitle{Interest Extraction via LLM Prompting.}
For each item $i$ with textual metadata $(\mathbf{t}_i, \mathbf{d}_i)$ and optional visual content $\mathbf{v}_i$ (processed via CMSA, see \Cref{sec:cmsa}), we prompt an LLM $\mathcal{M}$ to extract intent descriptors using a structured CoT template:
\begin{equation}
    \mathbf{z}_i = \mathcal{M}(\text{Prompt}_{\text{DCIM}}(\mathbf{t}_i, \mathbf{d}_i, \mathbf{v}_i^{\text{text}})),
\end{equation}
where $\text{Prompt}_{\text{DCIM}}(\cdot)$ guides the LLM through: (1) surface analysis of explicit item attributes, (2) inference of latent usage or purchase motivations implied by the item, and (3) synthesis of interpretable descriptor tags with confidence scores. Here $\mathbf{v}_i^{\text{text}}$ denotes the textual description of the visual content produced by CMSA.
\paratitle{Descriptor-Enhanced Text Representation.}
The mined descriptors $\mathbf{z}_i = \{z_i^{(1)}, z_i^{(2)}, \ldots, z_i^{(J)}\}$ are concatenated with the metadata and CMSA visual text to form an enriched textual input:
\begin{equation}
    \mathbf{x}_i^{\text{text}} = \text{Concat}(\mathbf{t}_i, \mathbf{d}_i, \mathbf{v}_i^{\text{text}}, \mathbf{z}_i).
\end{equation}
Unlike shallow item text, $\mathbf{x}_i^{\text{text}}$ encodes explicit attributes, visual textual evidence, and item-side intent descriptors. CMSA then fuses this text-side representation with the projected image embedding before RQ-VAE quantization (\Cref{sec:cmsa}). This encourages items with similar usage motivations and visual evidence to be mapped to nearby regions in the SID space, enabling the generative model to learn intent-aware item relationships.

\subsection{Cross-Modal Semantic Augmentation (CMSA)}\label{sec:cmsa}

To reduce modality mismatch before quantization, we propose \textbf{Cross-Modal Semantic Augmentation (CMSA)} (see Figure~\ref{fig_framework}), which enriches the item representation through two complementary visual paths before SID construction. The first path textualizes images with a VLM so visual semantics can be consumed by the text encoder and DCIM. The second path embeds the original image with a vision encoder, projects it to the SID embedding dimension, and fuses it with the text-side representation before RQ-VAE. CMSA is intentionally compatible with the MiniOneRec-style tokenizer: the RQ-VAE architecture and downstream generative backbone remain unchanged, while the pre-quantization item representation is enriched.

\paratitle{Motivation.}
In multimodal recommendation, items are associated with both textual attributes (\eg title, description) and visual content (\eg product images). A caption path provides interpretable visual semantics, but it may omit fine-grained cues such as texture, shape, or layout. A visual embedding path preserves continuous visual features, but by itself may be less interpretable to an LLM-based descriptor miner. CMSA therefore uses both paths: visual text supports intent mining, while visual embeddings preserve image-specific information for quantization.

\paratitle{Path 1: VLM-Based Visual-to-Text Semanticization.}
For each item $i$ with visual content $\mathbf{v}_i$, we employ a Vision-Language Model $\mathcal{V}$ to generate a rich textual description that captures the semantic content of the image:
\begin{equation}
    \mathbf{v}_i^{\text{text}} = \mathcal{V}(\mathbf{v}_i, \text{Prompt}_{\text{align}}),
\end{equation}
where $\text{Prompt}_{\text{align}}$ instructs the VLM to describe recommendation-relevant visual attributes, such as aesthetic style, color scheme, usage scenario, and lifestyle signals. We concatenate this visual text with item metadata:
\begin{equation}
    \tilde{\mathbf{t}}_i = \text{Concat}(\mathbf{t}_i,\; \mathbf{d}_i,\; \mathbf{v}_i^{\text{text}}),
\end{equation}
where $\tilde{\mathbf{t}}_i$ serves as the input to DCIM (\Cref{sec:dcim}). The caption path therefore makes visual semantics available to the LLM descriptor miner.

\paratitle{Path 2: Visual Embedding Preservation.}
In parallel, we encode the original image with a vision encoder $f_{\text{vis}}(\cdot)$ and project it to the same dimensionality as the text-side representation:
\begin{equation}
    \mathbf{e}_i^{\text{vis}} = W_v f_{\text{vis}}(\mathbf{v}_i),
\end{equation}
where $W_v$ is a learnable projection matrix. This path preserves continuous visual information that may not be fully captured by the caption.

\paratitle{Dual-Path Fusion before RQ-VAE.}
After DCIM produces item-side descriptors $\mathbf{z}_i$, we encode the enriched textual representation and fuse it with the visual embedding:
\begin{equation}
    \mathbf{e}_i^{\text{text}} = f_{\text{emb}}(\mathbf{x}_i^{\text{text}}),
\end{equation}
\begin{equation}
    \mathbf{e}_i^{\text{mm}} = W_f [\mathbf{e}_i^{\text{text}}; \mathbf{e}_i^{\text{vis}}],
\end{equation}
where $f_{\text{emb}}(\cdot)$ denotes Qwen3-Embedding-4B, $[\cdot;\cdot]$ denotes concatenation, and $W_f$ projects the fused vector to the RQ-VAE input dimension. The resulting multimodal embedding $\mathbf{e}_i^{\text{mm}}$ is then quantized by the same RQ-VAE configuration used in MiniOneRec-style SID construction. Thus, CMSA preserves both interpretable visual semantics and continuous visual features without changing the tokenizer architecture.

\subsection{Quality-Aware Reinforcement Mechanism (QARM)}\label{sec:qarm}

To improve downstream generation over the constructed SIDs, we propose the \textbf{Quality-Aware Reinforcement Mechanism (QARM)}, as depicted in the right part of Figure~\ref{fig_framework}. QARM uses supervised fine-tuning (SFT) for initial alignment, precomputes descriptor-quality labels, and then applies reinforcement learning (RL) with a relevance-gated quality reward. The RL stage optimizes the generative policy over SID sequences; it does not modify the VLM/LLM descriptor generator or retrain the RQ-VAE codebook.

\paratitle{Stage 1: Supervised Fine-Tuning.}
In the SFT stage, we train the generative model to predict target SID sequences given user histories represented by SIDs constructed from CMSA/DCIM-enriched item representations. The training objective minimizes the negative log-likelihood:
\begin{equation}
    \mathcal{L}_{\text{SFT}} = -\sum_{(\mathbf{X},\mathbf{Y}) \in \mathcal{D}} \sum_{h=1}^{H} \log p_\theta(y_h | \mathbf{X}, y_1, \ldots, y_{h-1}),
\end{equation}
where $\mathcal{D}$ is the training dataset, $\mathbf{X}$ is the user-history SID sequence built from intent-enriched item representations, and $\mathbf{Y}$ is the target SID sequence. This stage aligns the model with collaborative patterns over the constructed SID vocabulary; the DCIM descriptors affect SFT through the item SIDs rather than as online text inputs.

\paratitle{Pre-RL QARM Binary Quality Labeling.}
Before RL training, we employ a lightweight LLM-based binary classifier to assess the quality of mined descriptors. Specifically, we use a Qwen-series model in a zero-shot manner to classify each descriptor $z_i^{(j)}$ into a binary quality label:
\begin{equation}
    l_i^{(j)} = \text{LLM}_{\text{cls}}(z_i^{(j)}) \in \{0, 1\},
\end{equation}
where $l_i^{(j)} = 1$ indicates a \textit{positive} descriptor (specific, actionable, and grounded in item content) and $l_i^{(j)} = 0$ indicates a \textit{negative} descriptor (vague, generic, or hallucinated). When an item has multiple descriptors, we aggregate descriptor labels into an item-level quality score:
\begin{equation}
    q_i = \frac{\sum_{j=1}^{J_i} w_i^{(j)} l_i^{(j)}}{\sum_{j=1}^{J_i} w_i^{(j)}},
\end{equation}
where $w_i^{(j)}$ is the confidence score returned by the DCIM prompt. If confidence is unavailable, we set all $w_i^{(j)}=1$. This score is used only as an auxiliary semantic-quality signal during policy optimization. As a minimal reliability check, two annotators independently labeled 500 randomly sampled descriptor--item pairs using the same criterion as QARM (specific, actionable, and grounded in item evidence); their substantial agreement (Cohen's $\kappa=0.762$) suggests that descriptor quality is a recognizable annotation target, though not a stand-alone proxy for personalized relevance.

\paratitle{Quality-Aware Reward Function.}
The reward function combines standard recommendation rewards with a relevance-gated semantic-quality bonus. Given a generated SID sequence $y$ and the ground-truth target item $i^*$, the exact-match reward is:
\begin{equation}
    r_{\text{exact}}(y) = \mathbb{1}[\text{SID}(y) = \text{SID}(i^*)].
\end{equation}
Following prior RL baselines for SID generation, we also include SID-format rewards that provide partial credit when a candidate is syntactically valid or shares a correct prefix with the target SID:
\begin{equation}
    r_{\text{format}}(y) = \lambda_{\text{valid}}\mathbb{1}[y \in \mathcal{S}] + \lambda_{\text{pref}}\frac{\text{LCP}(y,\text{SID}(i^*))}{H},
\end{equation}
where $\mathcal{S}$ is the valid SID vocabulary, $\text{LCP}(\cdot)$ is the longest common prefix length, and $H$ is the SID length. These format rewards are not unique to our method; they are used as standard RL stabilizers for sparse SID generation.

The QARM-specific term is the semantic-quality bonus. To avoid rewarding an irrelevant item only because its mined descriptor is fluent, the bonus is gated by exact target-item decoding:
\begin{equation}
    r_{\text{quality}}(y) = \mathbb{1}[\hat{i}(y)=i^*] \cdot q_{\hat{i}(y)},
\end{equation}
where $\hat{i}(y)$ denotes the item decoded from SID sequence $y$, and $q_{\hat{i}(y)}$ is the aggregated QARM quality score. The final reward is:
\begin{equation}
    r(y) = r_{\text{exact}}(y) + r_{\text{format}}(y) + \alpha \cdot r_{\text{quality}}(y),
\end{equation}
where $\alpha=0.5$ balances semantic quality with recommendation correctness. Thus, QARM refines the reward among correctly decoded target-item predictions; it does not provide positive reward to semantically rich but incorrect items.

\paratitle{Stage 2: GRPO Optimization with QARM Rewards.}
We employ Group Relative Policy Optimization (GRPO)~\cite{shao2024deepseekmath} for efficient RL training. For each query, we sample a group of $G$ interest-enriched SID candidates (\eg $\text{item}_1, \ldots, \text{item}_G$) and compute group-normalized advantages using the quality-aware rewards:
\begin{equation}
    \hat{A}_j = \frac{r_j - \text{mean}(\{r_j\}_{j=1}^G)}{\text{std}(\{r_j\}_{j=1}^G)+\epsilon},
\end{equation}
where $\epsilon=10^{-6}$ prevents division by zero when all sampled candidates receive identical sparse rewards. The final RL objective includes a KL-divergence regularization term to prevent policy drift from the SFT initialization:
\begin{equation}
    J(\theta) = \mathbb{E}_{y \sim \pi_\theta} \left[ r(y) - \beta D_{\text{KL}}(\pi_\theta \| \pi_{\text{ref}}) \right],
\end{equation}
where $\pi_{\text{ref}}$ is the SFT reference policy and $\beta$ is the regularization coefficient. This closes the downstream learning loop: CMSA and DCIM construct intent-enriched SIDs, while QARM trains the generative policy to prefer correct SID predictions whose item descriptors are also specific and grounded.

\begin{table*}[t]
\centering
\caption{Overall performance comparison on three Amazon Product Reviews datasets. Bold indicates the best performance, and underline indicates the second best. $\Delta$ denotes the relative improvement of DeepInterestGR over the best baseline.}
\label{tab:main}
\resizebox{\textwidth}{!}{
\begin{tabular}{l|cccc|cccc|cccc}
\toprule
\multirow{2}{*}{\textbf{Method}} & \multicolumn{4}{c|}{\textbf{Beauty}} & \multicolumn{4}{c|}{\textbf{Sports}} & \multicolumn{4}{c}{\textbf{Instruments}} \\
& HR@5 & HR@10 & N@5 & N@10 & HR@5 & HR@10 & N@5 & N@10 & HR@5 & HR@10 & N@5 & N@10 \\
\midrule
\rowcolor{gray!20} \multicolumn{13}{l}{\textit{Traditional Sequential Models}} \\
GRU4Rec & 0.0312 & 0.0518 & 0.0189 & 0.0256 & 0.0198 & 0.0324 & 0.0118 & 0.0161 & 0.0285 & 0.0467 & 0.0172 & 0.0231 \\
Caser & 0.0287 & 0.0483 & 0.0171 & 0.0234 & 0.0175 & 0.0291 & 0.0103 & 0.0142 & 0.0261 & 0.0432 & 0.0156 & 0.0212 \\
HGN & 0.0335 & 0.0549 & 0.0201 & 0.0271 & 0.0212 & 0.0348 & 0.0126 & 0.0172 & 0.0302 & 0.0495 & 0.0183 & 0.0246 \\
\midrule
\rowcolor{gray!20} \multicolumn{13}{l}{\textit{Transformer-based Models}} \\
SASRec & 0.0398 & 0.0645 & 0.0241 & 0.0323 & 0.0248 & 0.0401 & 0.0149 & 0.0201 & 0.0361 & 0.0583 & 0.0218 & 0.0292 \\
BERT4Rec & 0.0421 & 0.0682 & 0.0256 & 0.0342 & 0.0267 & 0.0432 & 0.0161 & 0.0217 & 0.0385 & 0.0621 & 0.0233 & 0.0312 \\
S$^3$-Rec & 0.0445 & 0.0718 & 0.0271 & 0.0361 & 0.0283 & 0.0457 & 0.0171 & 0.0230 & 0.0407 & 0.0656 & 0.0247 & 0.0330 \\
FDSA & 0.0432 & 0.0698 & 0.0263 & 0.0351 & 0.0274 & 0.0443 & 0.0165 & 0.0223 & 0.0394 & 0.0636 & 0.0239 & 0.0320 \\
\midrule
\rowcolor{gray!20} \multicolumn{13}{l}{\textit{Generative \& LLM-based Models}} \\
TIGER & 0.0487 & 0.0763 & 0.0302 & 0.0395 & 0.0312 & 0.0498 & 0.0192 & 0.0256 & 0.0445 & 0.0712 & 0.0273 & 0.0362 \\
LC-Rec & 0.0523 & 0.0821 & 0.0328 & 0.0428 & 0.0341 & 0.0543 & 0.0212 & 0.0281 & 0.0478 & 0.0765 & 0.0295 & 0.0390 \\
HSTU & 0.0578 & 0.0897 & 0.0368 & 0.0476 & 0.0385 & 0.0611 & 0.0241 & 0.0318 & 0.0532 & 0.0845 & 0.0335 & 0.0438 \\
MiniOneRec & \underline{0.0612} & \underline{0.0945} & \underline{0.0389} & \underline{0.0502} & \underline{0.0398} & \underline{0.0627} & \underline{0.0251} & \underline{0.0330} & \underline{0.0556} & \underline{0.0878} & \underline{0.0347} & \underline{0.0455} \\
BIGRec & 0.0534 & 0.0839 & 0.0336 & 0.0439 & 0.0349 & 0.0556 & 0.0218 & 0.0289 & 0.0487 & 0.0779 & 0.0302 & 0.0399 \\
D3 & 0.0498 & 0.0785 & 0.0312 & 0.0409 & 0.0325 & 0.0519 & 0.0202 & 0.0269 & 0.0456 & 0.0732 & 0.0282 & 0.0374 \\
S-DPO & 0.0589 & 0.0918 & 0.0372 & 0.0483 & 0.0385 & 0.0609 & 0.0242 & 0.0319 & 0.0537 & 0.0852 & 0.0334 & 0.0440 \\
\textbf{DeepInterestGR} & \textbf{0.0678} & \textbf{0.1032} & \textbf{0.0436} & \textbf{0.0558} & \textbf{0.0452} & \textbf{0.0703} & \textbf{0.0289} & \textbf{0.0376} & \textbf{0.0623} & \textbf{0.0972} & \textbf{0.0394} & \textbf{0.0513} \\
\midrule
$\Delta$ Improv. & +10.8\% & +9.2\% & +12.1\% & +11.2\% & +13.6\% & +12.1\% & +15.1\% & +13.9\% & +12.1\% & +10.7\% & +13.5\% & +12.7\% \\
\bottomrule
\end{tabular}
}
\end{table*}

\section{Experiments}\label{sec:exp}

In this section, we empirically evaluate the effectiveness of the proposed DeepInterestGR framework. We aim to answer the following research questions:
\begin{itemize}
    \item \textbf{RQ1}: How does DeepInterestGR perform compared to competitive baselines across multiple benchmarks?
    \item \textbf{RQ2}: What is the individual contribution of each core component (DCIM, CMSA, and QARM) to the overall performance?
    \item \textbf{RQ3}: How does CMSA-based dual-path visual enrichment improve performance over text-only representations?
    \item \textbf{RQ4}: How does the relevance-gated QARM reward compare to alternative RL reward strategies, and how critical is the QARM binary quality labeling?
\end{itemize}
Beyond aggregate ranking metrics, we include branch-level CMSA ablations and failure-oriented SID analysis to examine whether multimodal intent enrichment changes the generated token structure for visually similar but intent-different items, rather than only improving top-$K$ accuracy.

\subsection{Failure-Oriented Case Study on Amazon Sports}\label{sec:casestudy}

To illustrate why pre-quantization visual evidence and item-side intent descriptors matter, we present anonymized Amazon Sports examples in \Cref{tab:casestudy}. The examples are organized around two failure modes of surface-only SID construction: (1) products with similar titles but different usage intent, and (2) products with short text whose usage context is mainly revealed by visual evidence. The underlying item identifiers and reproduced case-study metadata are provided in the anonymized code repository; the product names and images in the paper are anonymized to preserve the failure pattern without exposing product identifiers.

\begin{table*}[t]
\centering
\footnotesize
\setlength{\tabcolsep}{5pt}
\renewcommand{\arraystretch}{1.08}
\newcommand{\casecell}[1]{\begin{minipage}[t]{\linewidth}\vspace{0pt}\raggedright #1\end{minipage}}
\newcommand{\intent}[1]{\textcolor{red!70!black}{#1}}
\newcommand{\sidshift}[2]{{\bfseries\textcolor{red!70!black}{\texttt{#1} $\Rightarrow$ \texttt{#2}}}}
\newcommand{\sidline}[2]{\par\smallskip{\centering\textbf{SID:} \sidshift{#1}{#2}\par}}
\caption{Failure-oriented Amazon Sports examples. Text-only SID construction can group products by surface tokens, while DeepInterestGR uses dual-path CMSA and DCIM to recover item-side usage intent from visual evidence; item identifiers are anonymized.}
\label{tab:casestudy}
\begin{tabularx}{\textwidth}{@{}>{\raggedright\arraybackslash}p{2.55cm}
                              >{\raggedright\arraybackslash}p{2.75cm}
                              >{\raggedright\arraybackslash}p{4.55cm}
                              >{\raggedright\arraybackslash}X@{}}
\toprule
\textbf{Item} & \textbf{Surface Text} & \textbf{Visual Evidence / Text-only Failure} & \textbf{Descriptor and SID Effect} \\
\midrule
\rowcolor{gray!12}
\multicolumn{4}{@{}p{\textwidth}@{}}{\textbf{Case 1: Similar title, different intent}. Similar soccer-ball titles can hide different user intents; visual cues help separate youth recreation from adult training.} \\
\midrule
\casecell{\includegraphics[width=2.15cm,keepaspectratio]{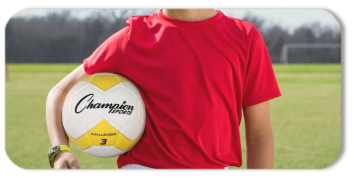}\newline
\textit{Youth / beginner ball}}
&
\casecell{\textbf{Title:} Classic Soccer Ball, Size 4, Blue/Yellow\newline
\textbf{Surface:} soccer ball, size 4, training, durable}
&
\casecell{Bright blue/yellow color; youth-oriented appearance; soft rubber texture; backyard or school-field scene.\newline
\textbf{Failure:} text-only SID may place it near generic training or match balls.}
&
\casecell{\intent{parent-purchased youth beginner soccer practice}; school/recreational play.\newline
\textbf{QARM:} 1; \textbf{Conf.:} 0.89\sidline{a1,b33,c192}{a1,b08,c144}}
\\
\cmidrule(lr){1-4}
\casecell{\includegraphics[width=2.15cm,keepaspectratio]{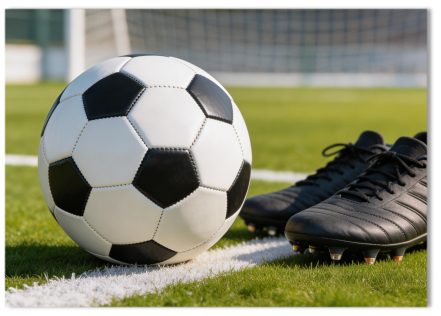}\newline
\textit{Adult / match ball}}
&
\casecell{\textbf{Title:} Classic Soccer Ball, Size 5, White/Black\newline
\textbf{Surface:} soccer ball, size 5, training, durable}
&
\casecell{Standard match-ball texture; stitched panels; grass/turf field; full-size goal and club-practice cues.\newline
\textbf{Failure:} surface text remains close to the youth ball, but usage intent differs.}
&
\casecell{\intent{competitive outdoor soccer training}; adult club practice; match preparation.\newline
\textbf{QARM:} 1; \textbf{Conf.:} 0.86\sidline{a1,b33,c197}{a1,b61,c209}}
\\
\midrule
\rowcolor{gray!12}
\multicolumn{4}{@{}p{\textwidth}@{}}{\textbf{Case 2: Short text, visual context needed}. When titles are short and underspecified, image context reveals indoor futsal control versus outdoor field practice.} \\
\midrule
\casecell{\includegraphics[width=2.15cm,keepaspectratio]{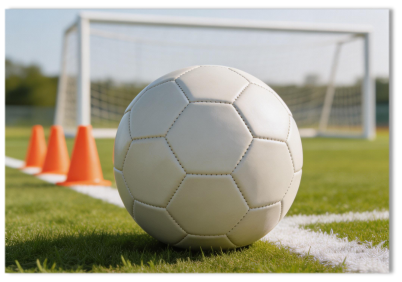}\newline
\textit{Indoor futsal ball}}
&
\casecell{\textbf{Title:} Soccer Training Ball, Size 4\newline
\textbf{Surface:} soccer training ball, size 4}
&
\casecell{Indoor court/gym-floor setting; compact size; futsal-style use; low-bounce or hard-court control cues.\newline
\textbf{Failure:} short title can collapse into ordinary soccer training balls.}
&
\casecell{\intent{indoor futsal low-bounce control training}; small-sided court play.\newline
\textbf{QARM:} 1; \textbf{Conf.:} 0.91\sidline{a1,b42,c083}{a1,b74,c031}}
\\
\cmidrule(lr){1-4}
\casecell{\includegraphics[width=2.15cm,keepaspectratio]{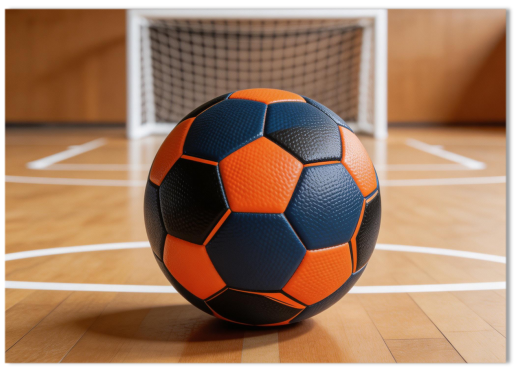}\newline
\textit{Outdoor training ball}}
&
\casecell{\textbf{Title:} Soccer Training Ball, Size 5\newline
\textbf{Surface:} soccer training ball, size 5}
&
\casecell{Grass/turf field; outdoor durability cues; water-resistant usage; team practice and full-size field context.\newline
\textbf{Failure:} surface text is close to the indoor futsal item, but context indicates outdoor match training.}
&
\casecell{\intent{outdoor field soccer practice}; team drills; grass/turf match training.\newline
\textbf{QARM:} 1; \textbf{Conf.:} 0.88\sidline{a1,b42,c091}{a1,b27,c186}}
\\
\bottomrule
\end{tabularx}
\end{table*}

\paratitle{Observations.}
\Cref{tab:casestudy} shows that the qualitative benefit of DeepInterestGR is not merely generating more fluent item text. In Case 1, two products share the same high-level category and similar title tokens, but the images imply different user intents: youth recreation versus adult competitive training. In Case 2, the product titles are short and underspecified, while the images and usage cues distinguish indoor futsal control from outdoor field practice. Dual-path CMSA exposes both caption-level visual semantics and continuous image features before quantization, and DCIM converts these signals into item-side intent descriptors. Qualitatively, this changes the generated SID structure: text-only SIDs tend to share two-token prefixes for visually similar soccer-ball products (\eg \texttt{a1,b33} for youth and adult training balls, and \texttt{a1,b42} for indoor and outdoor training balls). DeepInterestGR preserves the coarse soccer-ball root token \texttt{a1}, but separates the second-level prefix according to youth practice, match training, indoor futsal control, and outdoor field practice. This helps the SID represent why an item is useful, rather than only what surface category it belongs to.

\paratitle{Experimental Setup.}
We conduct experiments on three real-world public datasets from Amazon Product Reviews~\cite{mcauley2015amazon}: \textbf{Beauty}, \textbf{Sports and Outdoors} (Sports), and \textbf{Musical Instruments} (Instruments). Following prior work~\cite{rajput2023recommender,hou2023learningsid,zhou2020s3rec}, we apply the 5-core filtering protocol and the leave-last-out evaluation protocol. These categories were chosen before evaluation to cover appearance-driven goods (Beauty), activity/usage-driven goods (Sports), and specialized preference-driven goods (Instruments). The three datasets contain 22K--36K users, 10K--18K items, and 198K--296K interactions with sparsity around 0.0004--0.0008. We acknowledge that this is a category-level evaluation rather than a full-Amazon claim; extending to more categories is an important next step. Baselines span traditional sequential models (GRU4Rec, Caser, HGN), Transformer-based models (SASRec, BERT4Rec, S$^3$-Rec, FDSA), generative SID models (TIGER, LC-Rec, HSTU, MiniOneRec~\cite{kong2025minionerec}), and RL/preference-optimized variants (BIGRec, D3, S-DPO). We adopt \textbf{HR@K} and \textbf{NDCG@K} ($K \in \{5, 10\}$) as evaluation metrics with beam size 20. All experiments are conducted on NVIDIA A100 GPUs with \textbf{Qwen2.5-7B-Instruct} as the generative backbone and \textbf{Qwen3-Embedding-4B} as the item/descriptor text encoder. SFT uses learning rate $3 \times 10^{-4}$ for 3 epochs; GRPO uses $1 \times 10^{-5}$ for 2 epochs with KL coefficient 0.001, group size $G=8$, and advantage smoothing $\epsilon=10^{-6}$. To support reproducibility within the anonymized submission, the code repository contains the full hyperparameter grid, final configurations, encoder settings, prompts, random seeds, preprocessing scripts, and computing-resource details.

\subsection{Main Results (RQ1)}

\Cref{tab:main} presents the overall performance comparison of DeepInterestGR against all baselines on three Amazon Review datasets. We make the following observations:

\paratitle{DeepInterestGR achieves the best performance among compared baselines.}
DeepInterestGR outperforms all compared baselines across all three datasets and all evaluation metrics, achieving relative improvements of 9.2\%--15.1\% over the strongest baseline for each metric in \Cref{tab:main}. The strongest baseline is usually MiniOneRec, while S-DPO is the strongest RL/preference-optimized baseline. For example, on the Sports dataset, DeepInterestGR achieves HR@5 of 0.0452, surpassing MiniOneRec by 13.6\% and S-DPO by 17.4\%. These consistent gains across Beauty, Sports, and Musical Instruments suggest that intent-enriched SID construction is beneficial in the evaluated product domains, although we do not claim full-category generalization beyond these datasets.

\paratitle{Deep interest mining provides substantial gains over surface-level features.}
Comparing DeepInterestGR with MiniOneRec~\cite{kong2025minionerec} (same Qwen2.5-7B backbone and training protocol, but without CMSA/DCIM/QARM), we observe 10.8\%--13.6\% improvements in HR@5 across datasets. This gap quantifies the value of enriching SID inputs before quantization and adding relevance-gated semantic rewards: by capturing item-side usage motivations beyond surface-level product attributes, DCIM enriches the semantic content encoded into SIDs, enabling the generative model to learn more discriminative item representations.

\paratitle{Quality-aware reinforcement learning outperforms preference optimization baselines.}
S-DPO, which applies direct preference optimization for recommendation, represents the strongest RL/preference-based baseline in our comparison. DeepInterestGR surpasses S-DPO by 15.1\% in HR@5 on Beauty and 17.4\% on Sports. This result should be interpreted carefully: existing RL baselines already use SID-format or preference rewards to stabilize generation, whereas QARM adds a complementary semantic-quality term that is gated by exact target-item decoding. Thus the gain is not from using RL per se, but from adding item-descriptor quality as an auxiliary signal after correctness is enforced.

\paratitle{SID-based generative models perform strongly in the evaluated setting.}
Traditional sequential models (GRU4Rec, Caser, HGN) and Transformer-based models (SASRec, BERT4Rec, S$^3$-Rec) underperform the strongest SID-based generative baselines in \Cref{tab:main}. For instance, SASRec achieves HR@5 of 0.0398 on Beauty, while DeepInterestGR achieves 0.0678. This suggests that, in our evaluated setting, autoregressive modeling over semantic IDs can make effective use of the constructed item vocabulary, particularly when the SIDs are enriched with item-side intent descriptors from DCIM.

\subsection{Ablation Study (RQ2)}

To understand the contribution of each core component, we conduct ablation studies by systematically removing individual modules from the full DeepInterestGR framework. Results on the Beauty dataset are presented in \Cref{tab:ablation}.

\begin{table}[t]
\centering
\caption{Ablation study on the Beauty dataset. Each row removes one component from the full model.}
\label{tab:ablation}
\begin{tabular}{lcccc}
\toprule
\textbf{Variant} & \textbf{HR@5} & \textbf{HR@10} & \textbf{N@5} & \textbf{N@10} \\
\midrule
DeepInterestGR (Full) & \textbf{0.0678} & \textbf{0.1032} & \textbf{0.0436} & \textbf{0.0558} \\
\midrule
w/o DCIM & 0.0598 & 0.0921 & 0.0378 & 0.0489 \\
w/o CMSA (Text-Only) & 0.0641 & 0.0983 & 0.0412 & 0.0529 \\
w/o QARM Reward & 0.0635 & 0.0973 & 0.0408 & 0.0526 \\
w/o RL (SFT only) & 0.0567 & 0.0879 & 0.0358 & 0.0463 \\
\bottomrule
\end{tabular}
\end{table}

\paratitle{DCIM is the most critical novel component.}
Removing DCIM leads to the largest single-component performance drop: $-11.8\%$ in HR@5 and $-10.8\%$ in HR@10. This result indicates that descriptor mining is a key component of our framework. Without DCIM, the model degrades to encoding only surface-level textual features (title, description), losing item-side usage motivations that distinguish items with similar attributes. The magnitude of this drop supports our core hypothesis: instruction-tuned LLMs can help infer actionable item intent descriptors from metadata and visual textual evidence that are not explicit in shallow item text.

\paratitle{DCIM on surface-similar hard cases.}
The ablation above shows that removing DCIM substantially hurts performance, but it does not by itself distinguish DCIM from generic LLM rewriting. We therefore evaluate a surface-similar hard slice. For each dataset, we select test instances whose target item has high title-only TF-IDF similarity to another catalog item. This selection uses only surface title information and does not use DCIM descriptors, so the slice tests whether different item representations better handle lexically similar items rather than favoring DCIM.

\begin{table}[t]
\centering
\small
\caption{HR@10 on the surface-similar hard slice. The slice is selected using title-only TF-IDF similarity and does not use DCIM descriptors.}
\label{tab:similar_title_slice}
\begin{tabular}{lccc}
\toprule
Method & Beauty & Sports & Instruments \\
\midrule
w/o DCIM & 0.0832 & 0.0558 & 0.0770 \\
LLM Summarization & 0.0864 & 0.0582 & 0.0802 \\
LLM Paraphrasing & 0.0855 & 0.0575 & 0.0794 \\
DCIM & \textbf{0.0952} & \textbf{0.0645} & \textbf{0.0889} \\
\bottomrule
\end{tabular}
\end{table}

\Cref{tab:similar_title_slice} reports HR@10 on this slice. Generic summarization and paraphrasing improve over removing DCIM, indicating that better textual evidence is useful. However, DCIM consistently performs best on all three datasets, and its relative advantage over LLM summarization is larger on this hard slice than on the full test set. This suggests that DCIM contributes item-side intent-discriminative evidence beyond simply rewriting or summarizing item text, especially when surface titles are lexically similar.

\paratitle{Diagnostic scope.}
These ablations are intended to support a bounded mechanism claim rather than a broad claim about general-purpose LLM rewriting. All variants share the same backbone, SID construction, and training protocol, and the surface-similar slice is selected only from title-level TF-IDF similarity, independent of DCIM descriptors. Thus, the gains are best interpreted as evidence that intent-discriminative item evidence helps when surface metadata is ambiguous. Together with the CMSA and QARM ablations, the results narrow our conclusion to pre-quantization item-evidence enrichment and relevance-gated descriptor rewards.

\paratitle{CMSA reduces visual-text modality mismatch.}
Removing CMSA causes a $-5.5\%$ drop in HR@5 and $-4.7\%$ in HR@10. This suggests that dual-path visual enrichment provides meaningful gains beyond text-only item representations in this setting. We avoid interpreting CMSA as a new contrastive representation-learning method; the result supports the narrower claim that combining VLM-generated visual text with projected image embeddings before quantization improves SID-based recommendation.

\paratitle{QARM guides effective posterior optimization.}
Without the QARM quality-aware reward (replaced by standard exact-match and SID-format rewards), performance decreases by $-6.3\%$ in HR@5. This suggests that QARM provides useful auxiliary supervision beyond standard RL rewards. Because the semantic-quality term is gated by exact target-item decoding, it distinguishes semantically rich descriptors only among correct SID predictions rather than rewarding irrelevant but fluent item descriptions.

\paratitle{Reinforcement learning yields the largest overall improvement.}
The SFT-only variant shows the largest performance drop: $-16.4\%$ in HR@5 and $-14.8\%$ in HR@10. This highlights that the RL stage is important for SID generation in this setup. Importantly, QARM is not the only RL signal: the training also uses standard exact-match and SID-format rewards. QARM contributes the additional descriptor-quality term, whose value depends on DCIM producing specific and grounded descriptors.

\subsection{Analysis}

\subsubsection{CMSA Dual-Path Visual Enrichment Analysis (RQ3)}

We investigate whether CMSA-based dual-path visual enrichment further improves over text-only representations. To isolate the two visual paths, we compare four variants under the same DCIM, QARM, RQ-VAE, and generative backbone settings: (1) \textit{Text-Only} removes both visual paths; (2) \textit{Caption-only} keeps only the VLM-generated visual text path; (3) \textit{Image-embedding-only} keeps only the projected image-embedding path; and (4) \textit{Caption + Image Embedding} uses the full CMSA module.

\paratitle{CMSA: Fine-grained Visual Branch Ablation.}
\Cref{tab:multimodal} examines whether incorporating visual information through CMSA improves over text-only item representations. The Text-Only setting is aligned with the w/o CMSA row in \Cref{tab:ablation}: both remove the VLM-generated visual text path and the projected image-embedding path while keeping DCIM and QARM unchanged. Caption-only and Image-embedding-only isolate the two CMSA paths, while the full setting tests whether the two paths provide complementary evidence before SID quantization.

\begin{table}[t]
\centering
\caption{Fine-grained ablation of CMSA visual branches on the Beauty dataset. All variants use the same DCIM, QARM, RQ-VAE, and generative backbone settings; only the visual branch configuration changes.}
\label{tab:multimodal}
		\setlength{\tabcolsep}{4pt}  
\begin{tabular}{lcccc}
\toprule
\textbf{Setting} & \textbf{HR@5} & \textbf{HR@10} & \textbf{N@5} & \textbf{N@10} \\
\midrule
Text-Only & 0.0641 & 0.0983 & 0.0412 & 0.0529 \\
Caption-only & 0.0652 & 0.0995 & 0.0420 & 0.0537 \\
Image-embedding-only & 0.0648 & 0.0989 & 0.0417 & 0.0533 \\
Caption + Image Embedding & \textbf{0.0678} & \textbf{0.1032} & \textbf{0.0436} & \textbf{0.0558} \\
\midrule
$\Delta$ vs. Text-Only & +5.8\% & +5.0\% & +5.8\% & +5.5\% \\
\bottomrule
\end{tabular}
\end{table}

\paratitle{The two visual paths are complementary.}
Both individual visual paths improve over Text-Only. Caption-only improves HR@5 from 0.0641 to 0.0652 and NDCG@10 from 0.0529 to 0.0537, showing that VLM-generated visual text provides useful semantic evidence for descriptor mining and SID construction. Image-embedding-only also improves over Text-Only, reaching 0.0648 HR@5 and 0.0533 NDCG@10, indicating that continuous visual features preserve information not fully captured by metadata alone. The full Caption + Image Embedding setting achieves the best performance across all metrics, with +5.8\% HR@5 and +5.5\% NDCG@10 over Text-Only. This supports the design choice that caption-level semantics and projected image embeddings provide complementary pre-quantization visual evidence.

\subsubsection{QARM Reward Strategy and Quality Label Analysis (RQ4)}

We compare our relevance-gated QARM reward with alternative RL reward strategies and analyze the sensitivity of QARM binary quality labeling. Results are shown in \Cref{tab:reward}. All rows in the reward-strategy block use the same SFT initialization and GRPO training recipe; they differ only in the reward term used during RL.

\begin{table}[t]
\centering
\caption{Comparison of RL reward strategies and QARM quality label settings on the Beauty dataset. All settings start from the same SFT model.}
\label{tab:reward}
\begingroup
\small
\setlength{\tabcolsep}{1pt}
\begin{tabular}{@{}lcccc@{}}
\toprule
\textbf{Setting} & \textbf{HR@5} & \textbf{HR@10} & \textbf{N@5} & \textbf{N@10} \\
\midrule
\rowcolor{gray!20}
\multicolumn{5}{@{}l}{\textit{Reward Strategy Comparison}} \\
Standard SID reward (exact + format) & 0.0635 & 0.0973 & 0.0408 & 0.0526 \\
Collaborative & 0.0623 & 0.0954 & 0.0398 & 0.0513 \\
Prefix-Match & 0.0642 & 0.0981 & 0.0411 & 0.0528 \\
\midrule
\rowcolor{gray!20}
\multicolumn{5}{@{}l}{\textit{Descriptor-Quality Signal Variants}} \\
Uniform quality bonus & 0.0635 & 0.0973 & 0.0408 & 0.0526 \\
Random labels & 0.0598 & 0.0923 & 0.0381 & 0.0493 \\
Heuristic quality labels & 0.0644 & 0.0988 & 0.0413 & 0.0531 \\
\midrule
Relevance-gated QARM reward (Ours) & \textbf{0.0678} & \textbf{0.1032} & \textbf{0.0436} & \textbf{0.0558} \\
\bottomrule
\end{tabular}
\endgroup
\end{table}

\paratitle{QARM reward outperforms all alternatives.}
Among reward strategies, the standard SID reward combines exact target-item matching with SID-format rewards but does not use descriptor quality. Collaborative rewards, derived from user co-interaction patterns, provide a denser signal but can conflate popularity bias with genuine intent alignment. Prefix-match rewards partially address sparsity by rewarding partial SID matches, but still ignore the semantic quality of mined descriptors. Our relevance-gated QARM reward combines exact-match and SID-format rewards with QARM-labeled descriptor quality, providing semantic supervision only after exact target-item decoding is satisfied.

\paratitle{QARM label quality is critical for RL effectiveness.}
The QARM label quality analysis reveals a clear hierarchy: random labels hurt performance, suggesting that noisy quality signals can mislead policy optimization. Uniform quality bonuses collapse to the standard SID reward setting because every correctly decoded target item receives the same semantic bonus and no descriptor-quality distinction remains. Heuristic labels improve over random but fall short of LLM-based QARM, as they cannot reliably distinguish genuinely actionable descriptors from superficially specific but weakly grounded ones. Our LLM-based QARM classifier, operating in a zero-shot manner with structured prompts, uses the pretrained knowledge of Qwen-series models to assess descriptor specificity, actionability, and grounding.

\section{Conclusion}

We studied intent-enriched Semantic ID construction for multimodal generative recommendation and proposed \textbf{DeepInterestGR}. The framework combines three mechanisms: (1) \textbf{DCIM (Deep Contextual Interest Mining)}, which leverages LLM prompting to extract item-side intent descriptors from item metadata and visual textual evidence; (2) \textbf{CMSA (Cross-Modal Semantic Augmentation)}, a dual-path visual enrichment module that injects both VLM-generated visual text and projected image embeddings before quantization; and (3) \textbf{QARM (Quality-Aware Reinforcement Mechanism)}, which adds relevance-gated descriptor-quality rewards to standard RL training over SID sequences. Experiments on three Amazon Product Review categories show consistent improvements over competitive generative and RL/preference-based baselines. The conclusion is intentionally bounded: our evidence supports pre-quantization dual-path visual enrichment, item-side intent enrichment, and relevance-gated semantic rewards as useful mechanisms for SID-based generative recommendation, while broader category coverage and stronger multimodal representation learning remain important future work.

\bibliographystyle{IEEEtrans}
\bibliography{main}

\end{document}